\documentclass[amsmath,amssymb,aps, prl,
reprint,superscriptaddress,
]{revtex4-2}

\usepackage{graphicx,color,hyperref,%
fancyhdr,lastpage,physics,float}
 \newcommand\sbullet{\mathbin{\vcenter{\hbox{\scalebox{1.9}{$\bullet$}}}}}
 \newcommand\scirc{\mathbin{\vcenter{\hbox{\scalebox{1.9}{$\circ$}}}}}
\newcommand\Ket[1]{\big| {#1}\big\rangle}
\newcommand\Braket[3]{\big\langle{#1}\big| {#2}\big|{#3}\big\rangle}

\begin{document}
\preprint{APS/123-QED}

\title{Resonant fractional conductance through a 1D Wigner chain}

\author{Rose Davies}
\affiliation{
 School of Physics and Astronomy, University of Birmingham, Birmingham, B15 2TT, UK
}
\affiliation{
School of Computer Science \& Digital Technologies, Aston University, Birmingham, B4 7ET
}
\author{Igor V.  Lerner}
\affiliation{
 School of Physics and Astronomy, University of Birmingham, Birmingham, B15 2TT, UK
}
\author{Igor V. Yurkevich }
\affiliation{
School of Computer Science \& Digital Technologies, Aston University, Birmingham, B4 7ET
}


\date{\today}

\begin{abstract}
In recent experiments on conductance of one-dimensional (1D) channels in ultra-clean samples, a diverse set of plateaus were found at fractions of the quantum of conductance in zero magnetic field.   We consider a discrete model of strongly interacting electrons in a clean  1D system where the current between weak tunneling contacts is carried by fractionally charged solutions. While in the spinless case conductance remains unaffected by the interaction, as is typical for the  strongly interacting clean 1D systems,  we demonstrate that in the spinful case the peak conductance takes fractional values that   depend  on the filling factor of the 1D channel.
\end{abstract}
\synctex=1
\maketitle

Experiments on two-terminal conductance through one-dimensional (1D) systems contain many complex features despite the seeming simplicity of the reduced dimensionality. The most well known of these, along with the standard geometric quantization \cite{Imry:2008} of conductance in units of $2e^2/h $, is the 0.7 plateau \cite{07AnomalyPepper:96,07AnomalyMarcus:2002, 0.7review_2011}. Recent experiments have discovered a surprising new feature in the conductance -- additional plateaus occurring at fractional values of the conductance quantum at zero {(or very small)} magnetic field \cite{gul2018self,FractionPepper2019}.

The conductance through the prototypical 1D system, a clean Luttinger Liquid, is unaffected by interactions within the system since it is dominated in the dc limit by the contacts to reservoirs \cite{maslov1995stone,ponomarenko1995renormalization,safi1995transport}. It is only upon adding a scattering mechanism and more channels when fractional values of the conductance are expected \cite{Shavit2019Oreg,Aseev:2018}. While such phenomenologically introduced multi-particle backscattering was successfully utilized to reproduce one of the most prominent fractions of $2/5$ \cite{Shavit2019Oreg}, the even-denominator fractions have not been explained yet.

 {Typical samples in experiments \cite{gul2018self,FractionPepper2019} are ultra-clean an relatively short so that electron transport is ballistic. There is the experimental evidence \cite{PepperZigzag} of the formation of a zigzag Wigner crystal\cite{Matveev, Meyer2009} in precisely the same materials where the fractional conductance has been later discovered \cite{FractionPepper2019}.}

 {In this work we suggest  a discrete model of a clean 1D material with a strong electron-electron interaction where fractional charges, which can lead to the fractional conductance, arise due to incommensurability  {commensurability ?} of the Fermi wavelength and the effective lattice spacing. The fact that such a model results in the appearance of fractionally charged solitons \cite{Hubbard1978} has been established by symmetry arguments in a seminal work by Goldstone and Wilczeck \cite{GoldstoneFrac}. However, having fractional charges does not necessarily leads to the fractional quantization of conductance. We will show here that the latter arises only when additional channels, e.g.\ due to spin, are available to the electrons.}

 {The solitons  in question could arise due to the formation of the charge density wave (CDW) with a lattice constant incommensurate with the electrons Fermi wavelength.  In all these experiments, 1D constrictions can hold only a few electrons and a few superlattice periods, which makes imperative to build a finite-size model without going to the thermodynamic limit of the Luttinger liquid.}

 We consider the Hamiltonian of $N$ electrons hopping on a  lattice of $\ell $ sites with an infinite on-site repulsion (forbidden double occupancy) and a   next-neighbor repulsion:
\begin{align}
    H_0 &= \sum_{x=1}^{\ell -1} \left[-t\big(c^{\dagger}_{x}c_{x{+}1} + c^{\dagger}_{x{+}1}c_x \big) + U n_x\,n_{x{+}1}\right]\,.\label{spinlessHam}
\end{align}
 where $n_x=c^{\dagger}_xc_x$ is the on-site number operator. We assume the next-neighbor repulsion to be strong,
  \begin{align}\label{U gg t}
    U\gg t,
\end{align}
which effectively  projects out all states with two particles being on adjacent sites. The chain (\ref{spinlessHam}) is connected to the right and left reservoirs via the tunneling contacts with the couplings  $\Gamma_{{\mathrm{L,R}}}$.

We assume for definiteness that the number of sites $\ell $ is odd \footnote{Choosing $\ell $ to be even leads to similar results under an appropriate choice of the gate voltage.} and the gate voltage is such that the maximal number of states possible under condition (\ref{U gg t}), $\frac{1}{2}\qty(\ell {+}1)\equiv N{+}1$, is occupied making a string
\begin{align}\label{S}
S_{N{+}1}=\sbullet\,\scirc\,\sbullet\,\scirc\,\sbullet\cdots\,\scirc\,\sbullet
\end{align}
of alternate $N{+}1$ occupied ($\sbullet$)  and $N$ empty ($\scirc$) sites. Such a string corresponds to the ground state  of the system with $(N{+}1)$ particles. We consider the Hilbert subspace $\mathcal{H}_{N+1}$, where the high-energy states that contain adjacent occupied sites have been projected out under condition (\ref{U gg t}). In this subspace the ground state (\ref{S}) is unique and we denote it as
$\Ket{ {N{+}1}}$.

An important point is that under the same condition the ground state energy $E_N$ of the system with  $N$ particles can be very close to $E_{N{+}1}$ leading to the resonance conductance at low temperatures.

States with $N=\frac{1}{2}\qty(\ell -1)$ particles in the projected subspace $\mathcal{H}_N$, which are obtained from $\Ket{N{+1}}$, Eq.~\eqref{S}, by removing an electron from an odd (occupied) site, are made of two strings like that in Eq.~\eqref{S} separated a triplet of empty sites. Such states can be represented as
 \begin{subequations}\label{basis}
    \begin{align}\label{i=j}
        \Ket{i, i;N}\equiv \Ket{\otimes \,S_i\scirc\scirc\scirc\, S_{N-i}\,\otimes},\quad i=0,1\cdots N,
    \end{align}
 where the $2m{-}1$-long strings $S_m$ contain $m$ electrons with alternating occupied and empty sites  as defined in Eq.~(\ref{S}) for $m=N{+}1$. Here we represent left and right reservoirs attached to the chain as two additional sites at $x=0$ and $x=\ell {+}1$ depicted by crossed circles, which are never occupied by design. This is convenient to incorporate the  states obtained from $\Ket{N{+1}}$ by removal one of the edge electrons.   Assuming that a string $S_0$ is omitted together with an adjacent empty site (to keep the total number of sites unchanged), the boundary states   are
 \begin{align}\label{i=j=0N}
     \begin{aligned}
       \Ket{0, 0;N}&=\Ket{\otimes\scirc\scirc\,  S_{N}\,\otimes}  ,\,,  \\\Ket{N, N;N}&=\Ket{\otimes\,  S_{N}\,\scirc\scirc\,\otimes}\,.
     \end{aligned}
    \end{align}
     \end{subequations}
    Formally, the states in Eq.~\eqref{basis} are obtained by acting with the annihilation operator $c_x$ on any occupied site in $\Ket {N{+1}}$:
    \begin{align}\label{cN}
         \Ket{i, i;N} = c_{x}\Ket {N+1}\delta _{x,2i+1}\,.
    \end{align}
Further states in  subspace $\mathcal{H}_N$, obtained by acting with the hopping part of Hamiltonian (\ref{spinlessHam}) on states (\ref{basis}), can be represented  as configurations with two doublets of empty sites separating three strings,
\begin{subequations}\label{basis2}\begin{align}\label{i_not_j}
\Ket{i, j;N}=\Ket{ \otimes \,S_i\scirc\scirc  S_{j-i}\scirc\scirc  S_{N-j} \otimes},\quad i< j ,
\end{align}
including states with the boundary doubles:
 \begin{align}\label{0jN}
 \begin{aligned}
 \Ket{0, j;N}&=\Ket{\otimes\scirc\, S_{j}\scirc\scirc\, S_{N-j}\,\otimes} \,,\\
 \Ket{j, j;N}&=\Ket{\otimes\, S_{j}\scirc\scirc\, S_{N-j}\,\scirc\otimes} \,,\\
 \end{aligned}
\end{align}
 \end{subequations}

\begin{figure}
    \includegraphics[width = 0.95\columnwidth]{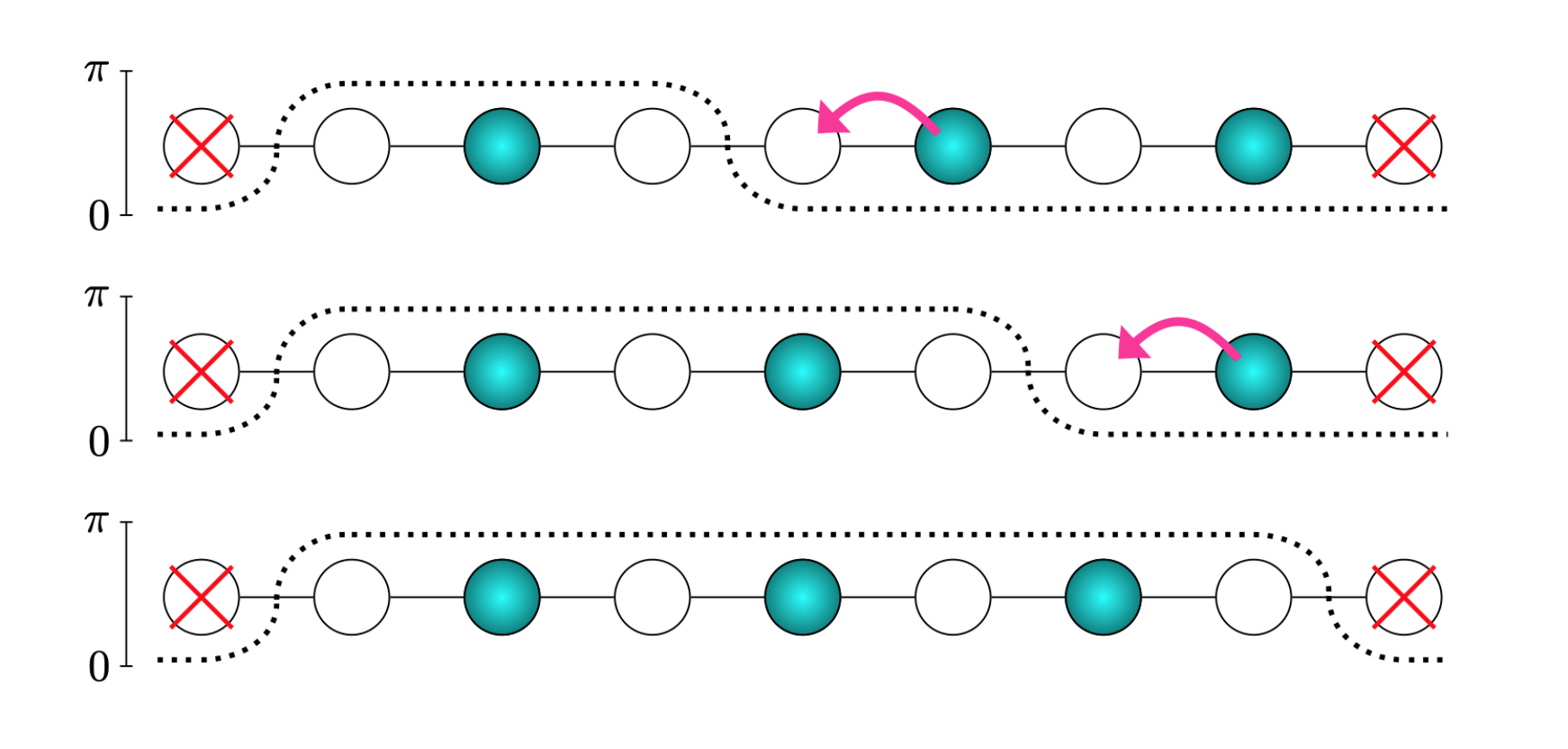}
    \caption{ The one-electron hopping from state $\Ket{0,1;3}$ to $\Ket{0,2;3}$ to $\Ket{0,3;3}$, Eq.~\eqref{0jN}, is equivalent to the motion of  the domain wall (kink), indicated by the dotted line. }
    \label{fig:HoppingShifts}
\end{figure}

In the states of Eqs.~(\ref{basis2}),   particles in  strings separated by two empty sites occupy either only even or only odd sites. The occupancy, $n_x=0$ or $1$, of any site can be represented as $n_x = \frac{1}{2}(\cos(\pi x {+} \phi_x)+1)$, with $\phi_x=0$ for occupied even (or empty odd) sites and $\phi_x=\pi$ for occupied odd (or empty even) sites. Hence, an electron hopping by one site can be  represented by the motion of the domain wall (kink) between $0$ and $\pi $ phases by two sites as illustrated in Fig.~\ref{fig:HoppingShifts}. In   the states of Eqs.~(\ref{basis}),  all electron occupied odd sites, and an electron hopping to an empty site is equivalent to the creation of kink -- anti-kink pair.

 As such pairs are created by removing a single electron from state $\Ket {N{+}1}$,   each kink carries the one-half electron charge, in agreement with the classical soliton picture of Goldstone and Wilczeck for polyacetylene \cite{GoldstoneFrac}. In a model similar to that under considerations, the existence of  such kinks has also been demonstrated numerically \cite{weiss2008finite}. We will show that, by itself, such a fractional charge does not lead to fractional conductance.

The states in Eqs.~\eqref{basis} and (\ref{basis2}), which are degenerate eigenstates of Hamiltonian $H_0$ in the absence of hopping, can be used as a basis for spanning any state $\Ket{\Psi_N}$
 in the projected subspace $\mathcal{H}_N$
 \begin{align}\label{N}
\Ket{\Psi_N}=\sum\limits_{0{\leqslant} i{\leqslant} j{\leqslant} N } \psi_{i, j{+}1}\,\Ket{i, j; N}\,.
\end{align}
To get the eigenvalue equation for $H_0$ within  $\mathcal{H}_N$, one needs to keep only the hopping terms acting on the ends of the strings, which results in
 \begin{align}\label{FF}
\varepsilon\,\psi_{ij}=-t \qty(\psi_{i+1, j}+\psi_{i-1, j}+\psi_{i,j{+}1}+\psi_{i,j-1} )\,.
\end{align}
This equation describes two free fermions of charge $-\frac{1}{2}e$ with positions $i$ and $j{+}1$ on the ficticious lattice of length $N{+}1$, with one being  on the left of the other,  $i\leqslant j$.
 The constraints on the indices in Eq.~\eqref{N}, $0{\leqslant} i{\leqslant} j{\leqslant} N$, can be accounted for by adding two boundary states, $i=-1$ and $j=N{+}1$, and imposing the boundary conditions
 \begin{align}\label{BC}
 \psi_{-1, j}=\psi_{i, N+1}=\psi_{ii}=0\,.
 \end{align}

  The solutions of Eq.~\eqref{FF} with the boundary conditions (\ref{BC}) are the Slater determinants of the standing waves
\begin{align}
\psi_{ij}(q_1, q_2)=\varphi_i(q_1)\,\varphi_j(q_2)-\varphi_i(q_2)\,\varphi_j(q_1)\,,
\end{align}
where
\begin{align}\label{varphi}
 \varphi_j(q)=\sqrt{\frac{2}{N{+}2}}\sin q(j{+}1)\,,\quad q=\frac{\pi n} {N{+}2}
\end{align}
with   $n=1,2,\cdots, N{+}1$.
The corresponding eigenenergies are
\begin{align}
\varepsilon(q_1, q_2)=-2t[\cos q_1+\cos q_2]\,.
\end{align}
The ground state, which we call $\Ket{N}$, is given by the lowest possible $q$, i.e.\
 \begin{align}\label{q1q2}
    q_1&=\frac{\pi }{N+2}\,,&q_2&=\frac{2\pi }{N+2}\,,
 \end{align}
  and its energy is $E_N=\varepsilon(q_1, 2q_1)$.

We consider the current through the system in the linear response regime under the conditions  when  only the states with $N{+}1=\frac{1}{2}(\ell {+1})$  and $N=\frac{1}{2}(\ell {-1})$ electrons in the chain are relevant as their  ground state energies are close. Then only the ground states, $\Ket{N+1}$ and $\Ket{N}$, contribute to the dimensionless conductance, which in the case of spinless fermions can  be written \cite{meir1992landauer} as
\begin{align}\label{T}
    g =\!  -\!\!\!\int\limits_{-\infty}^{\infty}\!\!\!  d\varepsilon\,T(\varepsilon)\,f'(\varepsilon-\mu)\,,\,\,
    T(\varepsilon)=4 \Gamma_{\mathrm{L}}\Gamma_{{\mathrm{R}}} \abs{G_{1\ell }(\varepsilon)}^2\!,
\end{align}
where $f(\varepsilon-\mu)$ is the electron Fermi distribution function and $G_{1\ell }(\varepsilon)$ is the retarded Green's function describing the propagation of effective excitations with energy $\varepsilon$ across an open chain
connected to the left, at $x=1$, and right, at $x=\ell $, reservoirs via the tunneling contacts with the couplings $\Gamma_{{\mathrm{L,R}}}$.

We shall use the Dyson equation,
  \begin{align}\label{Dyson}
G_{xx'}={\cal G}_{xx'}+\sum_{\alpha=1, \ell }{\cal G}_{x\alpha}\,\Sigma_{\alpha}\,G_{\alpha x'}\,,
\end{align}
to relate $G_{1\ell }(\varepsilon)$ to the Green's function ${\cal G}_{xx'}\qty(\varepsilon )$ of the isolated chain.
This equation for $G_{1\ell }$ is algebraic due to the locality of the self-energy: $\Sigma_{ 1}=-i\Gamma_{\mathrm{L}}$, $\Sigma_{ \ell }=-i\Gamma_{\mathrm{R}}$. Then the Green's function for a particle propagating across the system is found to be
\begin{align}\label{G}
     G_{1\ell } = \frac{\mathcal{G}_{1\ell }}{1 + i\mathcal{G}_{11}(\Gamma_{\mathrm{L}} + \Gamma_{\mathrm{R}}) + \Gamma_{\mathrm{L}}\Gamma_{\mathrm{R}}\big(\mathcal{G}_{1\ell }^2 - \mathcal{G}_{11}^2)}. 
\end{align}

The Green's function of the isolated chain, ${\cal G}_{x,x'}(\varepsilon)$,  is calculated assuming  infinitesimal coupling to the leads to ensure the thermal equilibrium. Keeping only the   states   $\Ket N$ and $\Ket{N{+}1}$, the retarded Green's function has a pole structure,
\begin{align}\label{Gcal}
{\cal G}_{x,x'}(\varepsilon)&=\frac{\rho_N+\rho_{N{+}1}}{\omega+i0}\,
\Braket{N}{c_x}{N{+}1}\Braket{N{+}1}{c^{\dagger}_{x'}}{N},\\
\omega&\equiv \varepsilon -\qty(E_N-E_{N+1})  ,\notag
\end{align}
where ${\rho_N}$ and $\rho_{N{+}1}$ are  canonical partition functions.

As follows from Eq.~\eqref{cN}, the  only states with $N$ particles that contribute to $\Braket{N}{c_x}{N{+}1}$ in Eq.~\eqref{Gcal} are $\Ket{i,i;N}$, Eq.~\eqref{basis}. Then we find from the expansion (\ref{N}) that
\begin{align}\label{reduction}
\Braket{N{+}1}{ c^{\dagger}_x}{N}=\psi_{i,i{+}1} \,, \quad x=2i+1\,.
\end{align}
Using the notations $x=2i{+}1, \, x'=2j{+}1$, we reduce the  Green's function (\ref{Gcal}) to
\begin{align}\label{z}
{\cal G}_{x x'}&=\frac{z_{i,j}}{\omega+i0},&
z_{i,j}\equiv ( \rho_N+\rho_{N{+}1}) \psi_{i, i{+}1}\psi_{j, j{+}1}\,.
\end{align}

The transmission coefficient then acquires the Breit-Wigner-Fano resonance form,
\begin{align}\label{Fano}
T(\omega)=\frac{4{\widetilde\Gamma_{\mathrm{L}}}{\widetilde\Gamma_{\mathrm{R}}} \omega^2}{\left[\omega^2-(1-s^2)\,{\widetilde\Gamma_{\mathrm{L}}} {\widetilde\Gamma_{\mathrm{R}}}\right]^2 +{\widetilde\Gamma}^2\,\omega^2}\,,
\end{align}
where the couplings to the reservoirs are renormalized by the Green's functions residues,
\begin{align}
{\widetilde\Gamma_{{\mathrm{L,R}}}}=z_{0,0}\,\Gamma_{{\mathrm{L,R}}}\,,\quad {\widetilde\Gamma}={\widetilde\Gamma_{\mathrm{L}}} +{\widetilde\Gamma_{\mathrm{R}}}\,,
\end{align}
with $s=z_{0,0}/z_{N,N}$ being the ratio of the residues. Taking into account that $\varphi_{N{+}1}=\pm\varphi_{0}$ and $\varphi_{N}=\pm\varphi_{1}$,  Eq.~\eqref{varphi}, the residues
 are equal to each other, i.e. $s=1$. The transmission coefficient, therefore, turns into standard Breit-Wigner formula,
\begin{align}
T(\omega)=\frac{4{\widetilde\Gamma_{\mathrm{L}}} {\widetilde\Gamma_{\mathrm{R}}}}{\omega^2+{\widetilde \Gamma}^2}\,.
\end{align}
For $\Gamma_{\mathrm{L}}=\Gamma_{\mathrm{R}}$ this peaks at $1$ at the resonance, $\omega\to 0$,  leading to the  universal value of conductance, $e^2/h$,  unaffected by the fractional character of the quasiparticles (kinks) inside the wire.  This is similar to the well-known result for the Luttinger liquid \cite{maslov1995stone,ponomarenko1995renormalization,safi1995transport}  where any internal interaction does not change, in the absence of backscattering, the universal conductance.

The generalization of the model (\ref{spinlessHam}) to a spinful case  drastically changes such a conclusion and results in fractional values of conductance.
Similar to the Luttinger liquid model, inclusion of the spin degrees of freedom in the model under considerations opens up an interaction channel not available in the spinless case. However, unlike the Luttinger liquid model there is no spin--charge separation in our setup. We show below that the addition of the spin degree of freedom suppresses the charge transport.

It turns out that in long constrictions  the conductance $g$ would be exponentially suppressed but in short constrictions, relevant for the experiments where the non-magnetic fractional conductance was discovered \cite{gul2018self,FractionPepper2019},  it takes fractional values that depend on the constriction length and filling factor.

To prove this claim, we generalize  Hamiltonian (\ref{spinlessHam}) to include spin $\sigma=\uparrow,\,\downarrow$:
\begin{align}\label{spin}
    c_x\to c_{x ,\sigma},\quad c^{\dagger}_{x}\to c^{\dagger}_{x,\sigma}\,,\quad
    n_x= n_{x,\uparrow} + n_{x,\downarrow}\,.
\end{align}
Each occupied state ($\sbullet$) in string (\ref{S}) and in the strings in Eqs.~(\ref{basis}) and  (\ref{basis2}) now acquires a spin index $\sigma$.

The spinful Green's function for an isolated chain, ${\cal G}_{x,x'}^{\sigma \sigma'}(\varepsilon)$,  is calculated along  the same route as the spinless one in Eqs.~(\ref{Gcal})--(\ref{z}). Now each configuration of strings  bears spin indices, which order cannot change as only the hopping part of $H_0$ affects the dynamics in the large-$U$ limit.

\begin{figure}
    \includegraphics[width = \columnwidth]{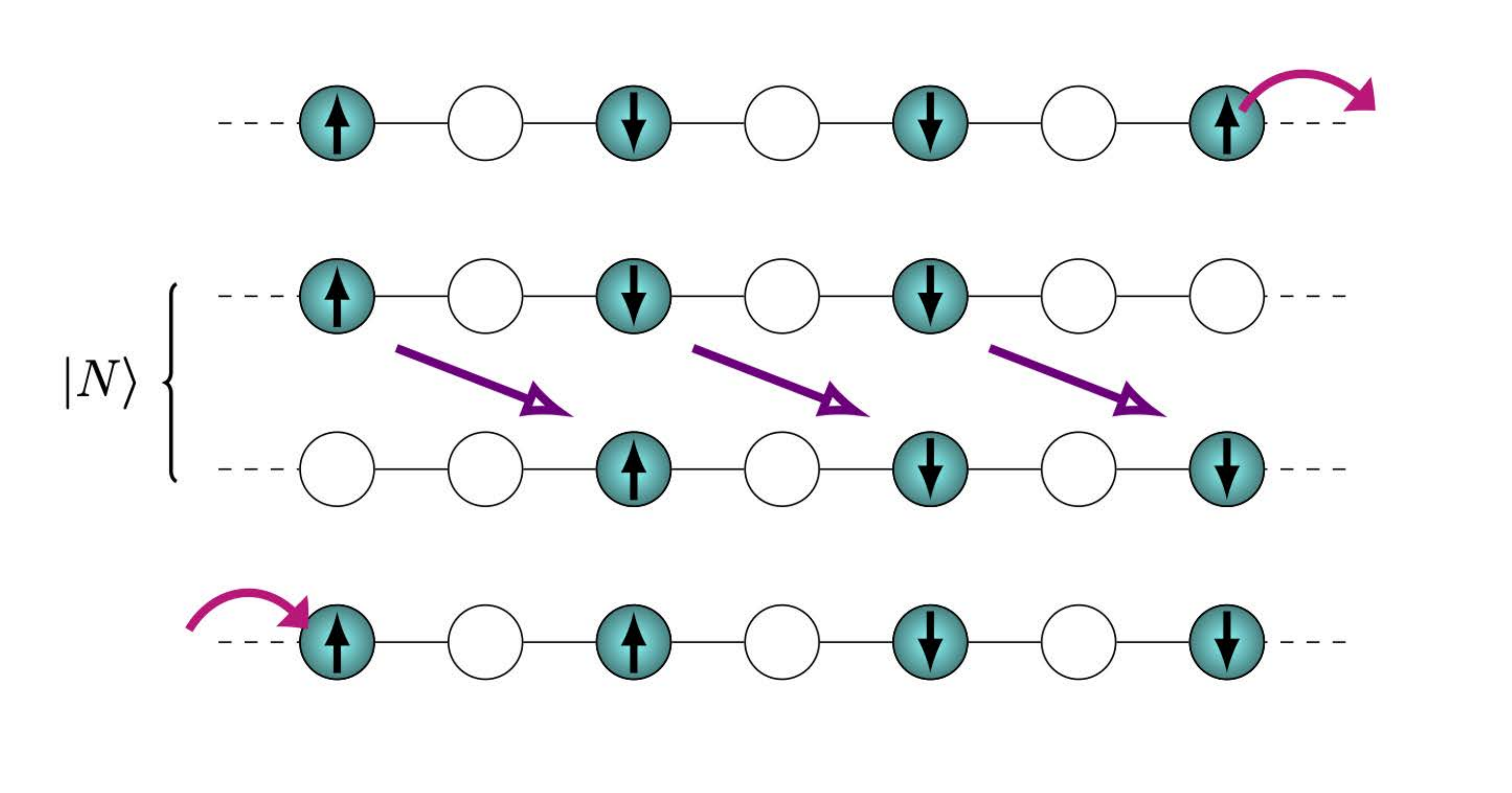}
    \caption{Demonstrating the orthogonality of a spin configuration under the hopping process. The dashes lines at either side of the chain represent the connection to the reservoirs.}
    \label{fig:Cotunneling}
\end{figure}

The spin  degrees of freedom in Green's function ${\cal G}_{x,x'}^{\sigma \sigma'}(\varepsilon)$ impose strong restriction on the matrix element $\Braket{N}{c_{x,\sigma}}{N{+}1}$ which should be substituted for $\Braket{N}{c_x}{N{+}1}$ in Eq.~\eqref{Gcal}: on top of the absence of the adjacent occupied states we should require that the spin configurations in $\Ket N$ and $\Ket{c_x}{N{+}1}$ are identical -- otherwise, these two states are orthogonal as illustrated in Fig.~\ref{fig:Cotunneling}.

As directions of spins in each string are arbitrary, the relative number of non-orthogonal configurations exponentially decreases with the length of the chain. Calculating this number with relatively straightforward combinatorics results in the following expression for the Green's function of an isolated chain in the spinful case, again using  the notations $x=2i{+}1, \, x'=2j{+}1$:
\begin{align}\label{Gspinful}
{\cal G}^{\sigma\sigma'}_{x x'}=\frac{z_{i,j}}{\omega+i0} A^{\sigma }_{ij} \delta_{\sigma\sigma'}\equiv {\cal G}^{\sigma }_{x x'}\,,
\end{align}
where the spin factor is
\begin{align}\label{Aij}
A^{\sigma }_{i,j}= 2^{N+i-j-1}\,,
\end{align}
and the residues $z_{i,j}$ are given by Eq.~\eqref{Gcal}. (Although $\rho_N$ and $\rho_{N{+1}}$ there include trace over spin configurations, it is not relevant for what follows).

The full Green's function $G_{1\ell }$, which enters  expression (\ref{T}) for conductance, is given by Eq.~\eqref{G} where ${\cal G}^{\sigma }_{x x'}\equiv {\cal G}^{\sigma\sigma'}_{x x'}\delta_{\sigma\sigma'}$ is substituted for ${\cal G} _{x x'}$.
To calculate the conductance, one needs only two components of it:
\begin{align}
{\cal G}^{\sigma}_{11}=  2^{N-1} {\cal G} _{11}\,,\quad {\cal G}^{\sigma}_{1\ell }= =\frac{1}{2}{\cal G} _{1\ell } \,.
\end{align}

Substituting this into Eq.~(\ref{Fano}) we find that $s=2^{-N} $ there with an additional factor of $2$ coming from the tracing over spin configurations, i.e.\ allowing for two spin channels. From this follows the main result: in the   zero temperature limit, at the resonant condition for equal coupling, the peak value of the transmission
\begin{align}
T_{\rm peak}=2^{-2N{+}1}\,
\end{align}
takes fractional values that should be observable in experiment on short constrictions.

\begin{figure}
    \includegraphics[width=0.95\columnwidth]{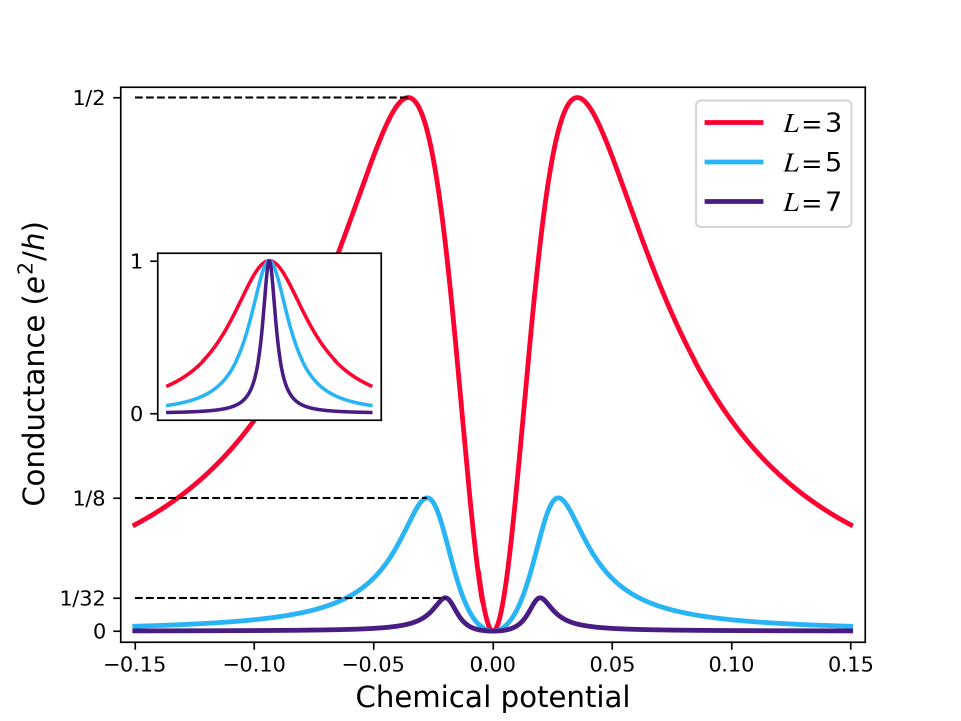}
    \caption{A graph showing the peaks in the spinful conductance and their reduction as the length of the system is increased as only transitions  that conserve spin configurations are   permitted. The energy has been set so that each length is plotted centered on its minimum energy}
    \label{fig:varyinglength}
\end{figure}

Figure \ref{fig:varyinglength} displays the spinful conductance at different lengths of chain.  The suppression of the conductance is caused by the increase in length of the system as the number of conducting spin configurations becomes a smaller part of the state space. This is in excellent correspondence to the experimental results on Germanium \cite{gul2018self}.

Recent experiments investigating the conductance of one-dimensional channels have unveiled a diverse range of plateaus occurring at fractions of the conductance quantum of in zero magnetic field. The adiabatic contacts in a clean interacting 1D system must guarantee insensitivity of the conductance to the microscopic details of interaction. Similar result is anticipated for the tunneling contacts under the resonant conditions. Our research demonstrates that this is true for spin-polarized electrons, where the peak resonant conductance matches the conductance quantum. However, in the case of spinful electrons experiencing strong electron-electron interaction, the peak conductance assumes fractional values dependent on the filling factor of the constriction and its length.

\begin{acknowledgements}
	We gratefully acknowledge support from EPSRC  under the grant EP/R029075/1 (IVL) and from the Leverhulme Trust under the grant  RPG-2019-317 (IVY).
\end{acknowledgements}

%

\end{document}